\documentclass[sigconf,natbib=false,screen]{acmart}

\pdfoutput=1

\usepackage{booktabs} 

\copyrightyear{2018} 
\acmYear{2018} 
\setcopyright{iw3c2w3}
\acmConference[SoCC '18]{SoCC '18: ACM Symposium on Cloud Computing    }{October 11--13, 2018}{Carlsbad, CA, USA}
\acmBooktitle{SoCC '18: ACM Symposium on Cloud Computing     (SoCC '18), October 11--13, 2018, Carlsbad, CA, USA}
\acmPrice{15.00}
\acmDOI{10.1145/3267809.3267822}
\acmISBN{978-1-4503-6011-1/18/10}

\usepackage{graphicx}
\usepackage{balance}  

\usepackage[utf8]{inputenc}
\usepackage[T1]{fontenc} 
\usepackage[american]{babel}

\usepackage{todonotes}
\usepackage{hyperref}
\usepackage{listings}
\usepackage{float}
\usepackage{stfloats}
\usepackage{graphicx}
\usepackage{caption}
\usepackage{subcaption}
\usepackage{xspace}
\usepackage{amsmath}
\usepackage{booktabs}
\usepackage{tabularx}
\usepackage{mathtools}
\usepackage[binary-units]{siunitx}
\usepackage{amssymb}
\usepackage{pifont}
\usepackage{balance}

\newcommand{\cmark}{\ding{51}}%
\newcommand{\xmark}{\ding{55}}%

\PassOptionsToPackage{
    sortcites,
    maxcitenames=2,
    maxbibnames=8,
  }{biblatex}
\usepackage{biblatex}
\bibliography{biblio}

\usepackage{xcolor}
\definecolor{ETHa}{RGB}{31,64,122}      
\definecolor{ETHb}{RGB}{72,90,44}       
\definecolor{ETHc}{RGB}{18,105,176}     
\definecolor{ETHd}{RGB}{114,121,28}     
\definecolor{ETHe}{RGB}{145,5,106}      
\definecolor{ETHf}{RGB}{111,111,100}    
\definecolor{ETHg}{RGB}{168,50,45}      
\definecolor{ETHh}{RGB}{0,122,150}      
\definecolor{ETHi}{RGB}{149,96,19}      

\PassOptionsToPackage{pdfpagelabels=false}{hyperref}
\usepackage{hyperref}

\newcommand{\bigquery}{{\it Google Big Query}\xspace}
\newcommand{\athena}{{\it Amazon Athena}\xspace}

\newcommand{\sss}{{\it Amazon S3}\xspace}
\newcommand{\qaas}{{\it QaaS}\xspace}
\newcommand{\qaasl}{query-as-a-service\xspace}
\newcommand{\Qaasl}{Query-as-a-service\xspace}
\newcommand{\sql}{SQL\xspace}

\newcommand{\qid}{{\texttt{query\_id}}\xspace}
\newcommand{\qset}{\texttt{query\_set}\xspace}

\makeatletter
\DeclareRobustCommand{\varname}[1]{\begingroup\newmcodes@\mathit{#1}\endgroup}

\definecolor{darkgreen}{rgb}{0.18,0.54,0.34}
\definecolor{codegreen}{rgb}{0,0.6,0}

\lstdefinelanguage{SSQL}[]{SQL}{%
    deletekeywords={count, sum},
    morekeywords={OR, WITH, AND, ELSE, LIKE, GROUP, PARTITION, OVER, UNNEST, TOP},
    morekeywords=[2]{COUNT, SUM, CARDINALITY, ARRAY_REMOVE, ARRAY_INTERSECT, ARRAY,
                     MULTISET, BIGINT, TINYINT},
    morekeywords=[3]{query_set, query_id},
}

\lstdefinestyle{sql} {
    language=SSQL,
    basicstyle=\ttfamily,
    keywordstyle=\color{blue},
    keywordstyle=[2]\color{orange!50!black},
    keywordstyle=[3]\it,
    commentstyle=\color{codegreen},
    stringstyle=\color{red!50!brown},
    columns=fullflexible,
    keepspaces=true,
    float=tbp,
}

\lstdefinestyle{ssql}{
    style=sql,
    basicstyle=\ttfamily\small,
}

\begin{document}
\title{Pay One, Get Hundreds for Free: \\
       Reducing Cloud Costs through Shared Query Execution}
\subtitle{Preprint}

\author{Renato Marroquín$^*$}
\email{marenato@inf.ethz.ch}
\affiliation{}

\author{Ingo Müller$^*$}
\email{ingo.mueller@inf.ethz.ch}
\orcid{0000-0001-8818-8324}
\affiliation{}

\author{Darko Makreshanski$^\dagger$}
\email{darko.makreshanski@oracle.com}
\affiliation{\hspace*{-9.1cm}\begin{minipage}[t]{\textwidth}\hspace*{\fill}$^*$Systems Group, Dept. of Computer Science, ETH Zürich\hspace*{\fill}$^\dagger$Oracle Labs\hspace*{\fill}\end{minipage}}

\author{Gustavo Alonso$^*$}
\email{alonso@inf.ethz.ch}
\affiliation{}

\renewcommand{\shortauthors}{R. Marroquín, I. Müller, D. Makreshanski, G. Alonso}

\hypersetup{%
    colorlinks=true,%
    urlcolor=ETHg,%
    linkcolor=ETHc,%
    citecolor=ETHd,%
    breaklinks=true,%
    pageanchor=true,%
}

\begin{abstract}

    Cloud-based data analysis is nowadays common practice because of the lower
    system management overhead as well as the pay-as-you-go pricing model. The
    pricing model, however, is not always suitable for query processing as heavy
    use results in high costs. For example, in query-as-a-service systems, where
    users are charged per processed byte, collections of queries accessing the
    same data frequently can become expensive. The problem is compounded by the
    limited options for the user to optimize query execution when using
    declarative interfaces such as SQL.  In this paper, we show how,
    without modifying existing systems and without the involvement of the cloud
    provider, it is possible to significantly reduce the overhead, and hence the
    cost, of query-as-a-service systems. Our approach is based on query
    rewriting so that multiple concurrent queries are combined into a single
    query. Our experiments show the aggregated amount of work done by the shared
    execution is smaller than in a query-at-a-time approach. Since queries are
    charged per byte processed, the cost of executing a group of queries is
    often the same as executing a single one of them. As an example, we
    demonstrate how the shared execution of the TPC-H benchmark is up to 100x
    and 16x cheaper in \athena and bigquery than using a query-at-a-time
    approach while achieving a higher throughput.


\end{abstract}

%
%

\ccsdesc[500]{Information systems~Query operators}
\ccsdesc[500]{Information systems~Structured Query Language}
\ccsdesc[500]{Information systems~Online analytical processing}
\ccsdesc[300]{Information systems~Relational parallel and distributed DBMSs}
\ccsdesc[300]{Information systems~Database performance evaluation}

\keywords{Data Warehouse, Shared Workload Execution, Query Processing, Cloud
    Computing, Serverless}

\maketitle

\section{Introduction}




\Qaasl (\qaas) enables users to query data already
hosted in the cloud without having to deploy extra infrastructure. Its pricing
model charges users only for the total number of bytes
processed by each query.  Applications accessing the same data set frequently
will become more expensive over time.  Examples of applications where sets
of queries will go repeatedly over the same data include search applications exploring a
solution space through parameter sweep queries to provide multiple alternative
answers (e.g., searching for airline tickets with multiple routes
\cite{Unterbrunner:2009:PPU:1687627.1687707}), reporting over different subsets
of the same data (e.g., maintaining BI dashboards
\cite{Wu:2014:CDV:2732951.2732964}), or what-if analysis.

Another appealing aspect of \qaas systems is the use of \sql for accessing and
managing data. Although, retrieving results is as easy as issuing \sql
statements, the possibilities for optimizing such systems are only at the \sql
level. Thus, users have almost no way to improve execution time further than
optimizing single query formulations and no obvious way to improve throughput
without directly increasing the monetary costs of executing queries.

The current pricing model from \qaasl systems, \athena and \bigquery, and the
limitations to optimize query execution motivate this work. In this context, we
extend the ongoing research on shared query execution to \qaasl systems by
exploiting sharing opportunities at the \sql level to reduce query
execution costs.  Existing work takes a rather invasive approach by modifying,
enhancing, or rewriting the query engine, which makes them not suitable for
\qaasl systems.

\begin{figure}[t]
    \includegraphics[width=0.4\textwidth]{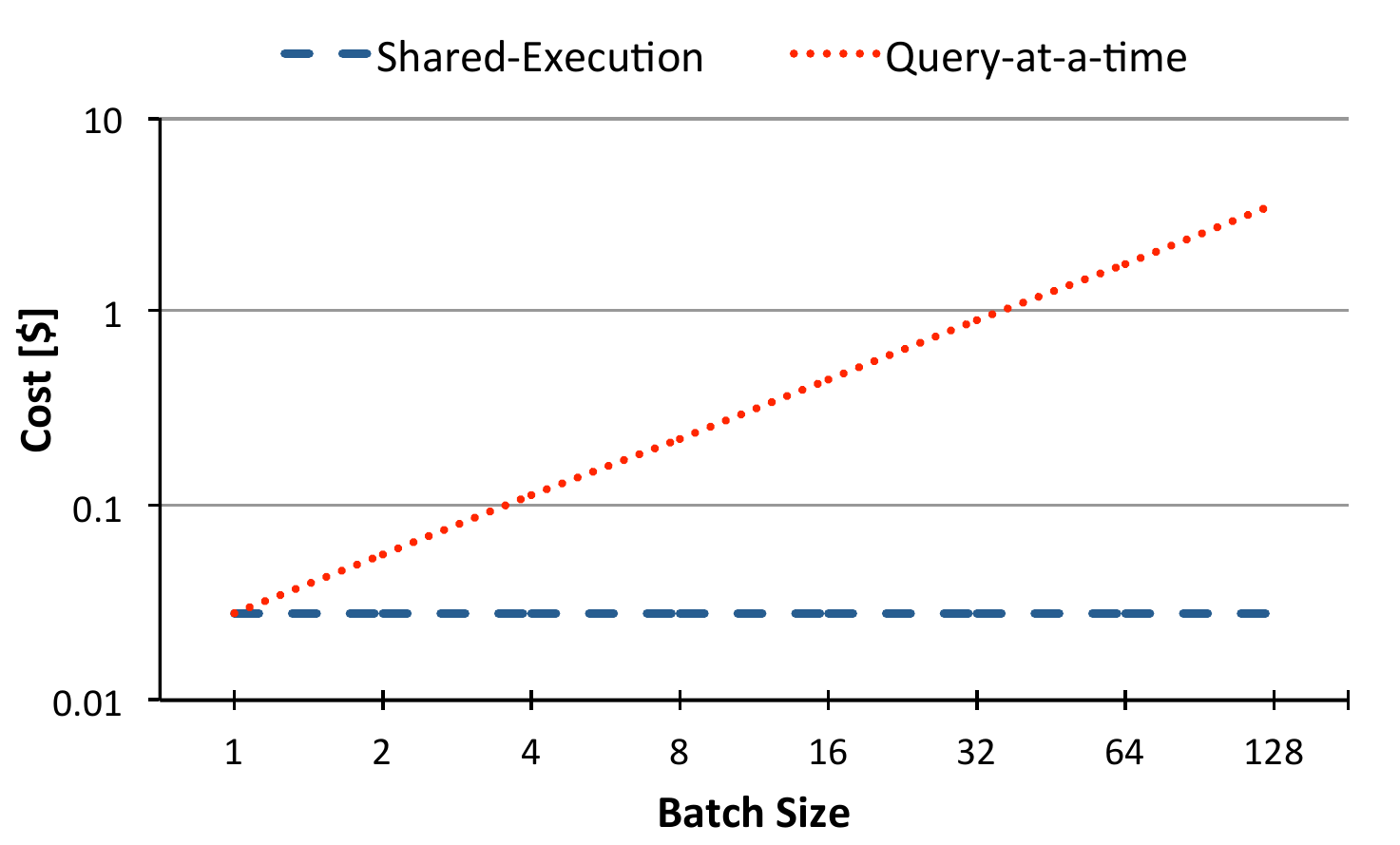}
    \centering
    \caption{TPC-H Q6 execution cost in \athena.}
    \label{fig:in:tpch6_qcost}
\end{figure}

In this paper, we show how to group and rewrite
\sql queries to be executed as a batch without modifying the underlying
system.
Queries are grouped and re-written as
part of an external middleware and the process does not require user input.
Thus, we trade off individual query latency for a throughput increase
while maintaining low execution costs.  This results in a smaller amount of work
to be done (i.e., data access) by the shared execution of multiple queries
compared to performing each query one at a time. In practice, the cost of executing a group
of queries is often the same as for executing a single query due to the current
\qaasl pricing model.  For example, Figure~\ref{fig:in:tpch6_qcost} shows the
execution cost in \athena of running up to 128 parameterized instances of TPC-H
    Query 6, i.e., each one requiring different subsets of data although all of
    them accessing the same base table.
Executing one
query after the other (following a query-at-a-time approach)
results in a very expensive workload. 
However, if we use a shared execution
approach and execute the queries together as a batch, we get a flat execution cost
regardless of the number of queries in the batch.  Even just a few queries
grouped together already provide significant savings. By grouping
128 queries together, we can increase
the throughput of this query by over 66x without increasing execution
cost over running a single query.



The main contributions of this paper are: 1) we enable 
cloud based \qaasl systems to perform shared execution without having to
re-engineer the underlying engine; 2) we present how relational operators
can be rewritten at the \sql level to support sharing by using a nested
representation of which tuple is of interest to a query; 3) we analyze the impact
of sharing for different operators and for more complex queries in terms of cost
and execution time on cloud based \qaasl systems; 4) we demonstrate the
potential of our approach with a TPC-H workload that we show executes 
up to two orders of magnitude
cheaper.

\section{Related Work}
\label{sec:rel-work}

Sharing computation among multiple concurrent queries was first explored in the
context of multi-query optimization (MQO)%
~\cite{Finkelstein:1982:CEA:582353.582400,Sellis:1988:MO:42201.42203}. The basic
idea consists of, given a set of queries, reducing the computational costs by performing 
shared expressions once, materializing them temporarily, and reusing them for solving the remainder of the queries.
Thus, the evaluation of common subexpressions 
is carried out only once. This approach was later
extended to benefit from query result caches \cite{qrc}, materialized/cached
views \cite{mcv}, and intermediate query results \cite{iqr1, iqr2}. However,
MQO does not use all
sharing potential.

More recently, a new line of work has developed ways to exploit sharing
opportunities such as sharing disk or memory bandwidth among queries without
common subexpressions.  For example, StagedDB \cite{stageddb} and QPipe
\cite{Harizopoulos:2005:QSP:1066157.1066201} use a simultaneous pipelining
technique to share work among queries that arrive within a certain time window.
MonetDB \cite{Zukowski:2007:CSD:1325851.1325934} and CoScan \cite{CoScan2011}
use cooperative scans where queries are dynamically scheduled together to
reduce the aggregated amount of I/O operations.  IBM UDB
\cite{Lang:2007:IBM:1325851.1325999} performs dynamic scan group and adaptive
throttling of scan speeds to suit a set of concurrent queries. CJoin
\cite{Candea:2009:SPJ:1687627.1687659} uses an always-on plan of join operators
to execute the joins of all concurrent queries. IBM Blink
\cite{Raman:2008:CQP:1546682.1547130} and Crescando
\cite{Giannikis:2010:CRE:1807167.1807326} answer multiple queries in one table
scan sharing disk and main-memory bandwidth. Data\-Path
\cite{Arumugam:2010:DSD:1807167.1807224} uses a push-based instead of a
pull-based model for a data-centric query processing to facilitate sharing of
concurrent queries.  SharedDB \cite{Giannikis:2012:SKO:2168651.2168654}
achieves predictable performance for highly concurrent workloads by query
grouping and using a global query plan to execute them.  MQJoin \cite{mqjoin}
efficiently shares the join execution for hundreds of concurrent queries.
These approaches significanly improve performance and demonstrate the potential
of sharing in many common workloads. However, they require significant changes
to existing database engines, thereby limiting their applicability
if modifying an existing system is not an option.  

Similarly to \cite{Candea:2009:SPJ:1687627.1687659, Arumugam:2010:DSD:1807167.1807224, Giannikis:2012:SKO:2168651.2168654, mqjoin}, our approach focuses on enabling
work sharing at run-time using an operator-centric approach, i.e., each operator process
a group of queries, thus exploiting both work and data commonalities at each operator.
To accomplish this, we annotate
intermediate results to obtain a high level of sharing for queries without
common subexpressions. 
The main distinction from previous work is that
we achieve this high degree of sharing solely through \sql rewriting, i.e.,
without requiring either modifications to the underlying engine or vendor
support. 
The goal in this paper is to explore the extent to which shared execution
pays-off and whether it can be implemented atop black-box query processing
engines such as those found in the cloud. 
In a related thesis~\cite{Wolf2017}, we explore enabling on-premise database
systems to support shared workload execution for some operators.  The results
of this paper extend this preliminary work.




\section{Multi Query Execution}
\label{sec:sw}

In this section, we first give an intuitive overview of how shared query
execution works.  Then we formalize this approach in the \emph{query-data model}
and define the relational operators for the model.

\subsection{Opportunities for shared execution}
\label{sec:opps_sexecution}

Sharing opportunities can be exploited whenever multiple queries need to access
the same base relations.  For example, performing a query in a search engine for
flight tickets is translated into a set of parameterized queries that
translate into potentially hundreds of individual queries \cite{Giannikis:2010:CRE:1807167.1807326} to offer
multiple options to the user.
In this scenario, we
could use work sharing across multiple queries by 
creating a batch out of them and then executing the batch in
one go. This optimizes data access and shares common computation among queries
at the expense of potentially increasing latency for individual queries.

Let us consider as a simpler example the two queries from Listing~\ref{ls:sw:sp_queries}.  They both
join the \lstinline[style=sql]{employees} table with the
\lstinline[style=sql]{departments} table on \lstinline[style=sql]{dept_id}, but
have different predicates.  
The two queries do not have common subexpressions.  However, there may still be
a large overlap among the tuples processed by the different queries, both in the
input and in intermediate results.

\begin{lstlisting}[
     caption={Set of individual queries.},
     captionpos=b,
     label=ls:sw:sp_queries,
     style=ssql
     ]
-- Query 1
SELECT * FROM employees E JOIN departments D
ON E.dept_id = D.dept_id
WHERE E.age = ? AND D.city = ?
-- Query 2
SELECT * FROM employees E JOIN departments D
ON E.dept_id = D.dept_id
WHERE E.name = ? AND D.address = ?
\end{lstlisting}


Thus to exploit more sharing opportunities, a single shared access plan can be
generated where the scan operation selects the \emph{union} of the input of both
queries, a single join of the respective results is carried out, and a
postprocessing step is done to extract the respective end results for each
individual query.  The benefit is that tuples relevant for the two queries are
processed only once.  Even though the total amount of tuples is larger than in
any single query, it is potentially much lower than the sum of the tuples needed
for each query.  It is thus often less work to run a single large plan than many
smaller plans.  In order to make sharing work, tuples needed by the shared plan
are annotated with the queries they are relevant to. To do this correctly
relational operators need to be adapted.

\subsection{Data-query model}
\label{sec:dq_model}

Shared query plans can be formalized using the \emph{data-query model}
\cite{Giannikis:2012:SKO:2168651.2168654}.
The main idea is to enhance
the relational data model with an extra attribute that tracks for which queries
each tuple is relevant.  We distinguish two different ways to do this
annotation: with atomic query identifiers and with sets of query identifiers.

When using atomic query identifiers,
we extend a relation $R$ with schema $R(A_1, A_2, A_3, \dots, A_n)$
by an additional attribute \qid:
\begin{equation}
  R'(A_1, A_2, A_3,\dots, A_n, \text{\qid}),
\end{equation}
where a tuple with $q = \text{\qid}$ is relevant for query%
\footnote{Where appropriate, we treat a query~$q$
         synonymously with its identifier.}~$q$
and tuples relevant for several queries are replicated once for each of them.
Any part of a shared query plan followed by a selection on $\text{\qid} = q$
and projection to $R$ is thus equivalent to the query plan of that of query~$q$.

When using sets of query identifiers,
we extend a relation $R$ with schema $R(A_1, A_2, A_3, \dots, A_n)$
by an additional attribute \qset:
\begin{equation}
  R'(A_1, A_2, A_3,\dots, A_n, \text{\qset}),
\end{equation}
where a tuple with $q \in \text{\qid}$ is relevant for query~$q$ and tuples
relevant for several queries occur only once.  Again, any part of a shared query
plan with the appropriate selection and projection is equivalent to that of a
query~$q$.  Relations may also not include any additional attribute, in which
case all tuples are relevant to all queries.

\begin{table}[t]
  \centering
  \begin{tabular}{cccc}
    \toprule
    \textbf{row\_id} & \textbf{Name} & \textbf{Other attr.} & \textbf{\qid} \\
    \midrule
    1                & EUROPE        & \dots                & 3             \\
    1                & EUROPE        & \dots                & 4             \\
    1                & EUROPE        & \dots                & 5             \\
    2                & AMERICA       & \dots                & 2             \\
    2                & AMERICA       & \dots                & 3             \\
    \bottomrule
  \end{tabular}
  \caption{Relation with a \qid attribute.}
  \label{tbl:dq-model-qid}
\end{table}

\begin{table}[t]
  \centering
  \begin{tabular}{cccc}
    \toprule
    \textbf{row\_id} & \textbf{Name} & \textbf{Other attr.} & \textbf{\qset} \\
    \midrule
    1                & EUROPE        & \dots                & 3,4,5         \\
    2                & AMERICA       & \dots                & 2,3           \\
    \bottomrule
  \end{tabular}
  \caption{Relation with a \qset attribute.}
  \label{tbl:dq-model-qset}
\end{table}

Tables~\ref{tbl:dq-model-qid} and \ref{tbl:dq-model-qset} show the same relation
in the data-query model using \qid and \qset attributes, respectively.  In both
cases, Queries~3 to 5 ``see'' the tuple with \texttt{row\_id} 1 and Queries~2
and 3 ``see'' the tuple with \texttt{row\_id} 2.

In this work, we annotate tuples first using a \qset attribute
and switching to use a \qid for the final
postprocessing step.

\subsection{Shared operators}
\label{sec:design}

To enhance relational operators to work in the data-query model, they have to
preserve the invariant that the tuples annotated with $q$
as well as those without \qid or
\qset attribute are the tuples relevant for query~$q$.  Operators on relations
without annotations do not need to be modified.

\subsubsection{Shared scan operator}

We start with the scan operator.  We call a scan operator a selection operator
whose input is not yet annotated with query identifiers, which is the case for
base tables.  Let $R$ be such a relation and $\sigma^{q_i} : R \rightarrow
\{\top,\bot\}$ the predicates for the queries in the batch $Q = \{q_1, \ldots,
q_n\}$.  The shared scan operator then works as follows:
\begin{equation}
  \resizebox{.9\hsize}{!}{
      $\sigma^Q(R) = \big\{ (t_R, \{q_i : \sigma^{q_i}(t_R)  = \top \} ) \; | \; t_R \in R \; : \exists q_i : \sigma^{q_i}(t_R)  = \top \big\}$
    }
  \label{eqn:shared_scan}
\end{equation}
and the schema of $\sigma^Q(R)$ is that of $R$ extended by a \qset attribute.
The value of this attribute is the set of query identifiers whose selection
predicate $\sigma^{q_i}$ evaluates to $\top$ on a particular tuple and
$\sigma^Q(R)$ only contains tuples where this is the case for at least one
query.

A selection operator on a relation with annotated tuples can be defined by
replacing the conditions $\sigma^{q_i}$ with ${\sigma'}^{q_i} = \sigma^{q_i}
\wedge q_i \in \text{\qset}$ or ${\sigma}'^{q_i} = \sigma^{q_i} \wedge q_i =
\text{\qid}$ for set-valued and atomic annotations, respectively.  Intuitively,
a tuple is in the result of query $q_i$ if it satisfies $q_i$'s predicate
$\sigma^{q_i}$ \emph{and} was relevant to $q_i$ before the selection.

\subsubsection{Shared join operator}

For the join operator, only the case
where both inputs are annotated is interesting.
In the other cases, a regular join can be used,
treating the \qid or \qset attribute like any other attribute (if present).
Let $R$ and $S$ thus be two relations with \qset attributes,
$f_{\bowtie} : R \times S \rightarrow \{\top,\bot\}$ a join condition for $R$ and $S$,
and $Q$ defined as above.
A join on these two relations is then defined as follows:
{
\begin{equation}
  \begin{alignedat}{3}
        R & \bowtie^Q S = \big\{ \rlap{$(t_R, t_S, R.\text{\qset} \cap S.\text{\qset} ) \; |$} && \\
        & t_R \in R, \; t_S \in S \; : \; && f_{\bowtie}(t_R, t_S) = \top \text{~and~} \\
        &                                 && R.\text{\qset} \cap S.\text{\qset} \neq \emptyset \}
  \end{alignedat}
  \label{eqn:shared_join}
\end{equation}
}
and the schema of $R \bowtie^Q S$ is that of $R \bowtie S$
extended by a \qset attribute.
The value of this attribute is the intersection
of the same attribute in R and S, respectively,
of tuples that match the join condition
and the result consists of those joined tuples
where this intersection is not empty.

A shared join on relations with \qid or mixed \qset/\allowbreak \qid attributes
can be defined in a similar way.
If both relations have a \qid attribute,
then $f_{\bowtie}$ is simply replaced by
\mbox{$f'_{\bowtie} = f_{\bowtie} \wedge t_R.\text{\qid} = t_S.\text{\qid}$}.
If they have mixed \texttt{query\_\allowbreak set}/\qid attributes,
the comparison in $f'_{\bowtie}$
is $t_R.\text{\qid} \in\allowbreak t_S.\text{\qset}$ or $t_R.\text{\qset} \ni t_S.\text{\qid}$.
In all three possible cases, the result has a \qid attribute.

\subsubsection{Shared grouping operator}
\label{sec:shared_groupby_op}

The grouping operator with aggregation
is slightly different than the ones above.
Since it computes new tuples out of several tuples from the input
(namely from those in the same group),
different queries generally only share a tuple in the output
if they already shared \emph{all} tuples in the input
that were used to compute that resulting tuple.
Since this is rather unlikely (and difficult to detect),
we define the shared grouping operator
such that no tuples are shared in the output,
i.e., we define it such that it always produces atomic \qid annotations.

Specifically, let $R$ be a relation with a \qset attribute,
$G \subset R$ a set of grouping attributes,
$F_i$ a set of aggregation functions on $R$,
and the query set $Q$ defined as above.
The grouping operator is defined as follows:
\begin{equation}
  \prescript{}{G}{\Gamma}^Q_{F_1,\ldots,F_k}(R) = \prescript{}{G,\text{\qid}}{\Gamma}_{F_1,\ldots,F_k}(R')
  \label{eqn:shared_grouping}
\end{equation}
where
\begin{equation}
    \resizebox{.9\hsize}{!}{
        $R' = \{ (t_R,q) \;|\; (t_R, \text{\qset}) \in R, q \in \text{\qset} \; \},$
    }
  \label{eqn:unnest_operator}
\end{equation}
which has the same schema as $R$,
except that it has a \qid instead of the \qset attribute,
and $\Gamma$ is the regular grouping operator from relational algebra.
We thus simply unnest the query identifiers in the \qset attribute
and group by the resulting \qid attribute in addition to $G$.
The grouping operator on \qid attributes
obviously works the same way, without prior unnesting.

\subsubsection{Shared operators in terms of regular operators}

In order to be able to express the above shared operators in \sql, the following
observation is crucial: all of them can be expressed in terms of unmodified
relational operators if an unnest operator is available.  For example, the
shared scan from Equation~\ref{eqn:shared_scan} can be re-phrased as follows:
\begin{equation}
    \resizebox{.9\hsize}{!}{
        $\sigma^Q(R) = \sigma_{\qset \neq \emptyset}\left(
            \Pi_{R, \qset \rightarrow \{q_i : \sigma^{q_i}(t_R)  = \top \}}(R) \right)$
    }
\end{equation}
Similarly, the definition of the shared join in Equation~\ref{eqn:shared_join}
can be expressed as a regular join that ignores the annotations of the input
followed by an appropriate projection and a filter dealing with the annotations.
Finally, the grouping operator from Equation~\ref{eqn:shared_grouping} is
already expressed using the regular grouping operator together with the
unnesting operator known from nested relational algebra~\cite{Colby1989}, which
is equivalent to Equation~\ref{eqn:unnest_operator}.

\subsection{Shared query plans}
\label{sec:shared_plans}

Producing a shared execution plan out of a group of queries has been studied in
the past. For example, the {\it Shared Workload Optimization algorithm} (SWO),
proposed by Giannikis et. al.~\cite{Giannikis:2014:SWO:2732279.2732280}, takes
an entire workload and produces a globally shared access plan. Similar
approaches for generating a shared execution plan is applicable in
our setting, thus they are not studied in this paper.

\section{Multi Query Execution as \sql}
\label{sec:rewriting}

%
%
%

In this section, we show how to express shared query plans as \sql.  The fact
that this is at all possible is based on the observation that we can express
shared operators in terms of standard relational operators.  Thus, we first
describe how shared operators can be expressed and further optimized in \sql and
then explain how such global plan can be successfully executed in \qaasl
systems.

\subsection{Shared operators}
\label{sec:exp_shared_ops_sql}


In the following, we show what data type to choose for the \qid and \qset
attributes, how to express the shared operators using \sql constructs, and how
to optimize some of the computations to increase efficiency.

\subsubsection{Tuple annotations}

We store a single query identifier as the smallest integer type
that can hold the largest number of queries in a batch,
e.g., \lstinline[style=sql]{TINYINT} for batches with up to 255 queries.
We use this type directly for \qid attributes.

For \qset attributes, standard \sql offers several ways for set-valued
attributes: \lstinline[style=sql]{ARRAY} (SQL:99 and up),
\lstinline[style=sql]{MULTISET} (SQL:2003 and up), \lstinline[style=sql]{BIGINT}
interpreted as bitset (any version), and possibly more.  The question of which
of them can be used depends on which set operations are supported by each type.
We need (1) construction of sets from atomics for the scan, (2) test for
emptiness for the scan and the join, (3) intersection for the join, and (4)
unnesting for the grouping operator.  While the standard defines all four
operations on \lstinline[style=sql]{MULTISET}s, most systems implement them for
\lstinline[style=sql]{ARRAY}s instead.  We thus use \lstinline[style=sql]{ARRAY}
as the type for \qset attributes in this paper.  In a related
thesis~\cite{Wolf2017}, we have explored how far one can get using
\lstinline[style=sql]{BIGINT}.

\subsubsection{Shared scan operator}
\label{sec:shared_scan}

As discussed in the previous section,
a shared scan operator is equivalent
to a projection computing a \qset attribute
followed by a selection to remove empty \qset{}s.
We propose a first way to achieve that in \sql
and an optimization in Section~\ref{sec:indexed-predicate-evaluation}.

Listing~\ref{ls:sscan_example} shows an example.
For each of the predicates $\sigma^{q_i}$ of the queries in the batch,
we create one \lstinline[style=sql]{CASE WHEN} statement
returning the query identifier if the predicate is fulfilled and 0 otherwise.
We store the result of these expressions in an array,
of which we remove the entries with 0,
thus obtaining only the desired identifiers
for the set of queries for which the tuple is relevant.
Since we evaluate one predicate after the other,
we call this approach \emph{linear predicate evaluation}.

\begin{lstlisting}[
    caption={Example of shared scan using linear predicate evaluation.},
    captionpos=b,
    label=ls:sscan_example,
    style=ssql
    ]
SELECT *,
   ARRAY_REMOVE(
     ARRAY[
       CASE WHEN id > 35              THEN 1 ELSE 0 END,
       CASE WHEN id BETWEEN 10 AND 20 THEN 2 ELSE 0 END,
       CASE WHEN id < 51              THEN 3 ELSE 0 END,
       CASE WHEN id BETWEEN 40 AND 50 THEN 4 ELSE 0 END
     ], 0) AS query_set
FROM employees
WHERE
   (id > 35) OR (id BETWEEN 10 AND 20) OR
   (id < 51) OR (id BETWEEN 40 AND 50);
\end{lstlisting}

For the selection of empty \qset{}s, we do a small optimization:
Instead of testing the arrays for emptiness,
we ``push the filter through the projection''
by testing instead for the disjunction of all predicates
before the arrays are even computed.
With linear predicate evaluation,
this was almost always faster in our preliminary evaluations,
in particular when this allows the database engine
to use min-max pruning.

The expression for computing the \qset attribute
could also be performed using user-defined functions (UDF).
Their performance heavily depends on implementation details
of the different systems.
UDFs can be beneficial in a system where they are Just-in-Time compiled
while expressions are interpreted.
However, UDFs might as well have an overhead
due to a function call for each evaluated tuple,
or not be supported at all.
For instance,
\athena does not support UDFs
and \bigquery currently supports JavaScript UDFs
with certain limitations~\cite{bq_udfs_limits}.

\subsubsection{Shared join operator}

\begin{lstlisting}[
    caption={Example of a shared join.},
    captionpos=b,
    label=ls:sjoin_example,
    style=ssql
    ]
WITH R AS (...),        -- shared left subplan
S AS (...),             -- shared right subplan
sjoin_helper AS (       -- join and compute query_set
    SELECT
        R.A1, ..., R.An, S.A1, ..., S.Am,
        ARRAY_INTERSECT(
            R.query_set, S.query_set) AS query_set
    FROM R JOIN S ON R.key = S.key)
SELECT *                -- filter out irrelevant tuples
FROM sjoin_helper  
WHERE CARDINALITY(query_set) > 0
\end{lstlisting}

As discussed above, a shared join can be expressed by a regular join followed by
a projection and a selection.  This can be done in a relatively straight-forward
manner in \sql.  Listing~\ref{ls:sjoin_example} shows an example.  We express
the join as a \lstinline[style=sql]{JOIN ... ON}, but other syntaxes can be
used.  The approach also extends beyond the equality join from the example.  In
order to compute the \qset attribute of the result, we use the array
\lstinline[style=sql]{ARRAY_INTERSECT} function.  Finally, we remove irrelevant
tuples by testing for emptiness of the computed \qset attribute.

The operations on arrays used in this example are vendor-specific.
However, as discussed above, the standard does define equivalent operations
and many database vendors implement some similar functionality.
Note that by using a \lstinline[style=sql]{BIGINT} representation
for \qset and bitwise $\varname{and}$ for set intersection,
it is possible to reimplement approaches like MQJoin~\cite{mqjoin} in \sql~\cite{Wolf2017}.

\subsubsection{Shared grouping and other operators}

\begin{lstlisting}[
    caption={Example of a shared grouping.},
    captionpos=b,
    label=ls:sgroupb_example,
    style=ssql
    ]
WITH sscan_emp AS (...),        -- shared scan
unnested_sscan AS (             -- unnest query_set
    SELECT * FROM sscan_emp
    WHERE CARDINALITY(query_set) > 0
    CROSS JOIN UNNEST(query_set) AS t(query_id))
SELECT   query_id, dept_id, COUNT(id) 
FROM     unnested_sscan
GROUP BY query_id, dept_id;     -- shared group-by
\end{lstlisting}

As discussed in Section~\ref{sec:shared_groupby_op}, a shared grouping operator
can be expressed as an \mbox{unnesting} operator on the \qset attribute followed by a
regular grouping operator where the resulting \qid attribute is added to the
grouping attributes.  Listing~\ref{ls:sgroupb_example} shows the implementation
of an example query in \sql.  The join on
\lstinline[style=sql]{UNNEST(query_set)} \lstinline[style=sql]{AS t(query_id)} replicates every tuple
once for each element in \qset and calls that element \qid.  The final grouping
is then a regular \lstinline[style=sql]{GROUP BY} clause.

Note that the unnesting of query identifiers increases the
size of the result of a shared subplan to the total aggregated result size of
each individual query subplans, i.e., no tuples are shared anymore.  This is
intrinsic to grouping with aggregation where every query requires its own tuples
and not specific to implementing sharing in \sql. In spite of this, a shared
grouping operator is still useful because the grouping result is small
compared to the input and also because the unnesting operation can be
efficiently implemented without the need to materialize a very large
intermediate result.



In case the original queries have an \lstinline[style=sql]{ORDER BY} operator,
we just prepend the \qset attribute to the ordering attributes.  Even
\lstinline[style=sql]{LIMIT}/\lstinline[style=sql]{TOP} clauses for shared plans
can be expressed in \sql using windowing functions, i.e., using a
\lstinline[style=sql]{PARTITION BY query_id} clause and number the records
within the partition of each query to then filter by that number.  This approach
works (and is required) for both computed and non-computed attributes.




\subsection{Shared scan with indexed predicate evaluation}
\label{sec:indexed-predicate-evaluation}

Shared scans using linear predicate evaluation
allows to share disk bandwidth,
saves work in downstream operators,
and can be expressed in \sql.
However, it has the same computational complexity
as a query-at-a-time approach:
each tuple is checked against the predicates of all queries.
In Crescando, Unterbrunner et al.~\cite{Unterbrunner:2009:PPU:1687627.1687707}
propose to index the constants of predicates
of the form $c_{\varname{lower}} < \varname{attribute} < c_{\varname{upper}}$
in order to evaluate the batch of predicates faster.

At first sight, implementing such an index in \sql seems impossible.
Interestingly, we can build a tree of \emph{expressions}
to evaluate all predicates of a batch
using a number of comparisons
that is proportional to the logarithm of the number of queries.
Like a ``real'' index, this reduces the evaluation cost of predicates
to a lower complexity class.
We call this approach \emph{indexed predicate evaluation}.

Building such an expression tree works as follows%
\footnote{The procedure essentially corresponds to building an interval tree.}:
We take all predicates as intervals of two constants
annotated by the query they belong to.
The root of the tree is a \lstinline[style=sql]{CASE WHEN} statement
testing for $\varname{attribute} < m$,
where $m$ is the median of the distinct interval bounds.
Then, we split up predicate intervals containing $m$ in two
and recurse using the intervals smaller than $m$
to build the expression tree for the $\varname{true}$ case
and the constants greater than $m$ for the other case.
For each subtree, we track the interval of possible values
that an attribute can have if that subtree is evaluated at scan time.
The recursion ends when the entire interval of the subtree
coincides with the predicate intervals in that subtree.
In this case, we know exactly the queries
whose predicates match the current tuple,
so we return an array with their identifiers.

\begin{lstlisting}[
    caption={Expression tree for indexed predicate evaluation.},
    captionpos=b,
    label=ls:predidx_example,
    style=ssql
    ]
(CASE WHEN id <= 35 THEN
    CASE WHEN id < 10 THEN ARRAY[3]
    ELSE
        CASE WHEN id <= 20 THEN ARRAY[2,3]
        ELSE ARRAY[3] END
    END
ELSE
    CASE WHEN id <= 50 THEN
        CASE WHEN id < 40 THEN ARRAY[1,3]
        ELSE ARRAY[1,3,4] END
    ELSE
        CASE WHEN id < 51 THEN ARRAY[1,3]
        ELSE ARRAY[1] END
    END
END) AS query_set
\end{lstlisting}

Listing~\ref{ls:predidx_example} shows the expression tree
that computes the \qset attribute of the shared scan
from Listing~\ref{ls:sscan_example}.
The outermost \lstinline[style=sql]{CASE WHEN} statement
tests for \lstinline[style=sql]{id <= 35},
which is the median of the constants ${10, 20, 35, 40, 50, 51}$.
If the $\varname{true}$ case is taken, we know that $\varname{id} < 35$,
which excludes the interval
\lstinline[style=sql]{BETWEEN 40 and 50} of query~4,
but includes some interval of all other queries,
in particular, the one-sided interval
\lstinline[style=sql]{id < 51} of query~1.
In the $\varname{true}$ case of the outer-most expression,
the next test is \lstinline[style=sql]{id < 10}.
From the remaining queries, only query~3 can satisfy this condition
and it does so for all possible values (namely for any $\varname{id} < 10$).
Hence, recursion ends and \lstinline[style=sql]{ARRAY[3]} is returned.
The other subtrees are built analogously.

Indexed evaluation is applicable to many types of predicates.
First, it works for any predicate based on the total order of a domain.
This includes equality, open and closed intervals, and one-sided intervals.
It also includes strings,
even with \lstinline[style=sql]{LIKE} expressions
as long as there is no wildcard in the beginning of the constant.
Second, it works for disjunctive predicates as well.
We simply treat each term in the disjunction of a query
like we treat an entire query in the procedure explained above,
but return the same query identifier for all of these terms in the leaves.

Last but not least, we can use indexed evaluation
for predicates on several attributes.
In this respect, our approach to handle several attributes is more general than the indexes of Crescando.
We pick a first attribute and build the expression tree
for predicates on that attribute as explained above.
In the leaves of the tree, we cannot return query identifiers yet
because we did not evaluate the predicates on the other attributes.
Instead, we continue building an expression tree,
but using the other attributes.
We recurse until the previous stopping condition is met
or the remaining predicates cannot be indexed,
in which case we do linear predicate evaluation.

One downside of indexed predicate evaluation
is the increased length of the query string.
It increases with the number of queries
depending how much their predicates overlap.
The two systems on which we evaluate our approach
both have a limit on the query string of \SI{256}{\kibi\byte}.
However, we do not reach that limit for any of the TPC-H queries
with batches of up to 128 queries.

\subsection{Shared query plans}
\label{sec:shared_qplans}

The shared access plan is a DAG-structured query plan, which assumes an engine
capable of executing and producing multiple outputs from a query execution.
However, current \qaasl are closer
to traditional execution engines in which queries are executed following a
Volcano-style processing \cite{Graefe:1993:VOG:645478.757691}, i.e., queries are
executed as tree-structured query plans.  This means that although queries can
be expressed as a single global plan, such a DAG-structured plan cannot be
directly executed.

To support the execution of DAG-structured query plans,
we convert a DAG into a set of tree-structured plans,
each of which can be executed as a single query.
To that aim, we identify operators in the execution DAG
whose output is used by multiple other operators.
For each of these operators, we have two options:
either we duplicate the operator including the tree of operators that it uses
(recursively)
or we materialize its output such that it can be read several times.
Which of the two is better can be decided
by using a cost-based optimizer as studied
by \cite{Neumann2009, Finkelstein:1982:CEA:582353.582400}.
Building such an optimizer is out of the scope of this paper,
so we do not discuss this aspect further.

\section{Evaluation}
\label{sec:exp_env}

To assess the behaviour of shared execution,
we benchmark shared operators in isolation to understand
how sharing impacts monetary cost of the system and query runtime
and evaluate the end-to-end behaviour by implementing a complete TPC-H
query workload.

\subsection{Experimental environment}

\textbf{Systems under test.} We evaluate two mainstream \qaasl systems,
\athena and \bigquery.

\athena uses a {\it pay-per-processed-byte} pricing model.
It consist of a fixed price for every byte read from \sss (S3) disregarding
how computationally expensive a query is or the size of intermediate results.
Thus, the chosen storage format has an impact on the actual query execution
cost. If the underlying data is stored in a row-oriented format, then the cost
for accessing a single attribute is the same as accessing all attributes. On
the other hand, if a column-oriented format (Apache Parquet or Apache ORC)
is used, then only the accessed attributes are relevant for the cost.

\bigquery uses a {\it pay-per-processed-byte} pricing model
that consists of a fixed price for every byte in the columns used by a query.
This is somewhat independent of how much bytes are actually read---%
if a column is used by a query, the query is billed
as if the column was read in its entirety.
Furthermore, similarly to \athena,
the storage format impacts directly the query execution cost
in that using a row-oriented format means that all columns are always used.

\textbf{Setup.} For each system under test, we use the recommended storage
format for obtaining the best possible results both in terms of execution time
and cost: Apache Parquet compressed columnar format stored in \sss for \athena
and \bigquery's native uncompressed columnar format.

For both systems and all experiments, we use a single connection,
such that queries (or query batches) are executed consecutively.
Both systems support multiple concurrent connections.
However, in experiments not shown here, we always observed
an ideal or near-ideal throughput improvement.
Furthermore, parallel execution could be applied to all approaches
shown in this section.
The effects shown in our experiments thus indicate \emph{efficiency}
that applies both to sequential and parallel execution.

In all experiments, we show the median of three runs. We perform two additional
warm-up runs for \athena, but omit them for \bigquery, as they had no effect.  
Moreover, we measure execution time and monetary costs of executing queries. We
do not measure the post-processing step for separating each query results
because filtering them is trivial in terms of size
and complexity compared to solving an actual query.

\subsection{Microbenchmarks of shared operators}
\label{sec:exp:sscan}

We first evaluate shared operators in isolation
in order to understand how various parameters
like the number of queries grouped together and their selectivities
influence their performance.
Due to space constraints,
we only show the results of the scan operator,
which---due to the pricing model---%
is the most relevant for monetary costs.

\subsubsection{Shared scan performance}
\label{sec:exp:sscan:impact_sel}

We use selection-only queries to observe how
the amount of data read and processed affects running time and monetary costs.
We use indexed predicate evaluation right away,
but quantify the impact of this optimization below.
For this experiment, we extend the LINEITEM table of TPC-H
with a column consisting of densely increasing integers
and run batches of queries,
each with a single, random predicate of a fixed selectivity using only that column.
At scale factor 100,
this table requires \SI{21}{\gibi\byte} and \SI{84.3}{\gibi\byte}
in \athena and \bigquery, respectively.
We use a selectivity of 99\% instead of 100\%
in order to prevent \athena
from skipping entire blocks based on Parquet metadata.

\begin{figure*}[bt]
    \centering
    \begin{subfigure}{.5\textwidth}
      \centering
      \includegraphics[width=0.9\linewidth]{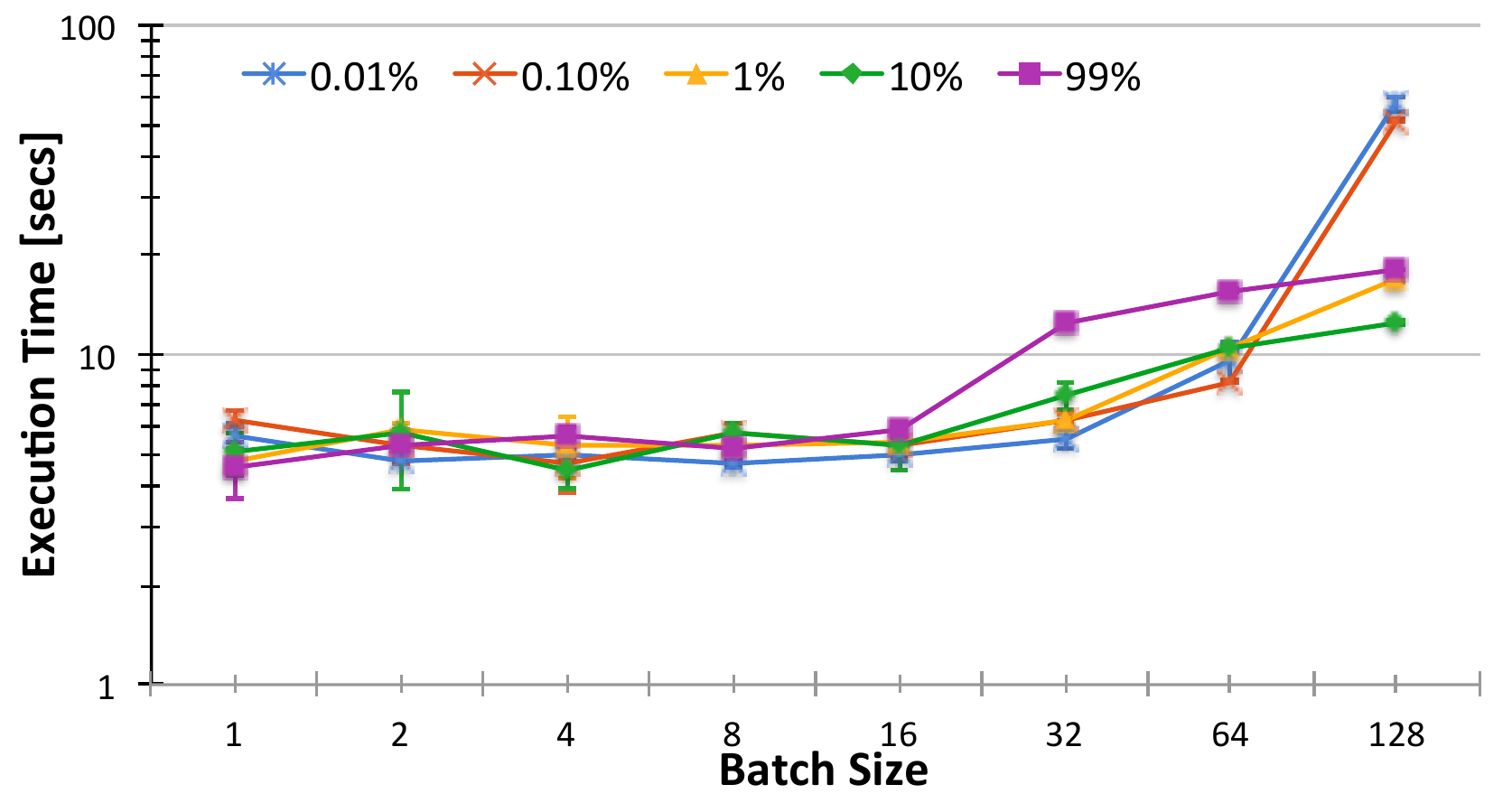}
      \caption{\athena}
      \label{fig:exp:sscans:exp1:sel:at}
    \end{subfigure}%
    \begin{subfigure}{.5\textwidth}
      \centering
      \includegraphics[width=0.9\linewidth]{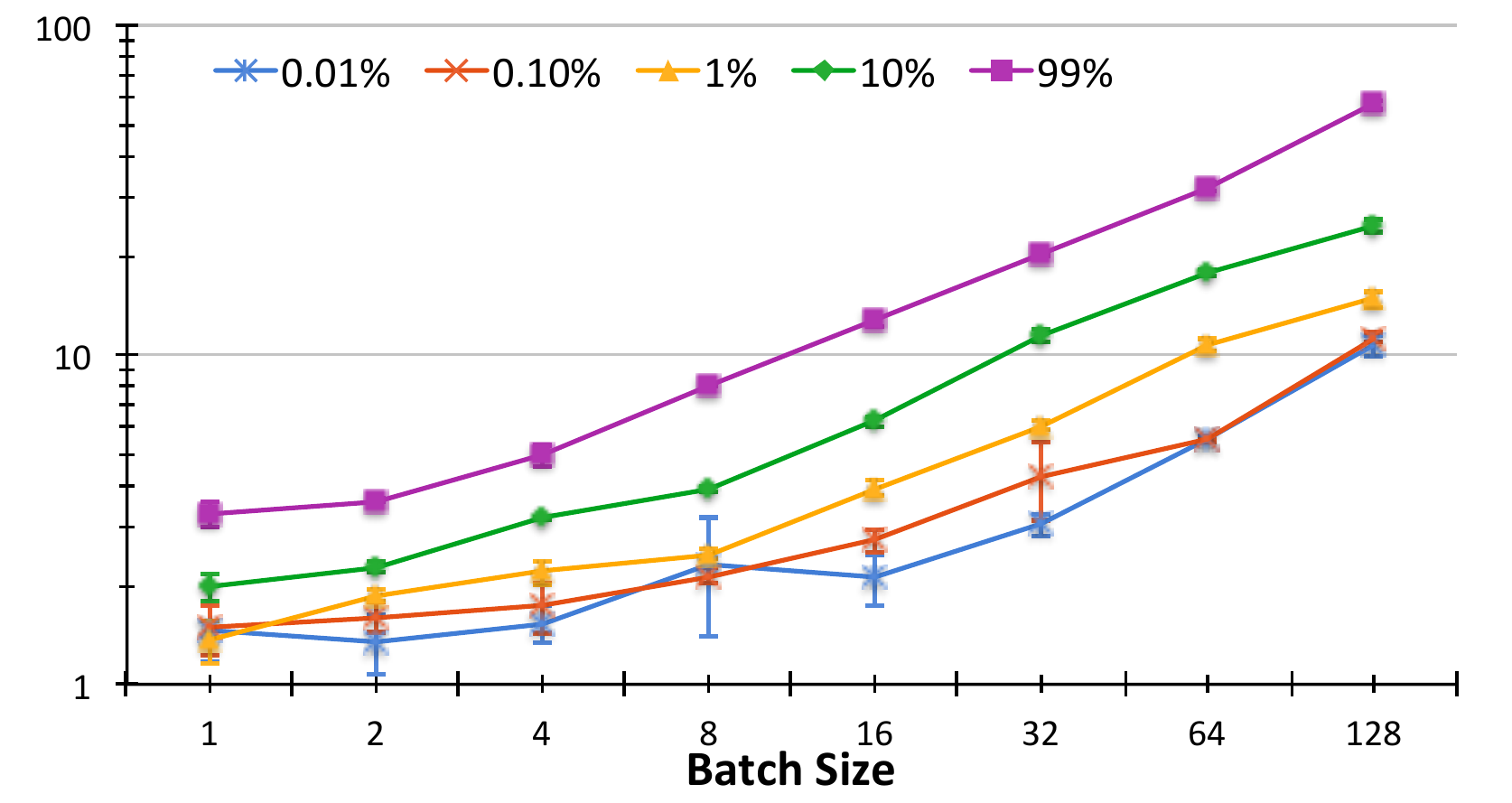}
      \caption{\bigquery}
      \label{fig:exp:sscans:exp1:sel:bq}
    \end{subfigure}
    \caption{Shared scan execution time for various selectivities.}
    \label{fig:exp:sscans:exp1:sel}
\end{figure*}

\textbf{Execution time.} Figure~\ref{fig:exp:sscans:exp1:sel:at} shows
the query execution times for \athena.
The execution time stays constant for batches of up to eight queries
and the running time is not affected by the selectivity.
This suggests that some constant costs such as job start-up
dominate the cost of the actual work.
In experiments not shown here, we tried with larger datasets
    but we observed the same effect.

With larger batches,
the running time increases
because (1) data volume and (2) computational complexity increase:
First, the more queries there are in a batch,
the greater their \emph{combined} selectivity
given a fixed \emph{per-query} selectivity.
Assuming $Q$ uncorrelated queries of selectivity $S$,
their combined selectivity is $(1 - (1 - S)^Q ) \cdot 100\% $.
This term approaches 100\% as the batch size increases
even if the per-query selectivity is small.
Second, each query in the batch may add computations for predicate evaluation,
even with predicate indexing,
which makes the scan compute-heavy for large batch sizes.
However, in most cases the running time increases
by a much lesser factor than the batch size,
suggesting an increase of efficiency due to a higher degree of sharing.

Similarly, the running time increases with the selectivity and the batch size,
which is particularly visible for the selectivity of 99\%.
Since the amount of data is virtually unaffected by the batch size,
this increase must be due to higher computational costs.
We explain this with the fact that,
for higher selectivities, each tuple is selected by more queries,
so the \qset attributes computed by the scan is larger.

Notice that the running time with a selectivity below 1\%
is almost 3x higher than that of selectivity 1\% for batches of 128 queries.
This is unexpected and does not fit the remaining observations.
We were able to reproduce a similar behaviour
in a local PrestoDB v0.170 installation,
but could not determine the root cause for the behaviour.
Further analysis and contacting support is required for this.

The fact that larger batch sizes
increase the execution time only by little or not at all
has a great effect on throughput:
If executing a batch of queries takes the same time as executing just one,
the throughput of a workload running in batches
is improved by the batch size
compared to the traditional query-at-a-time approach.
From the numbers show in Figure~\ref{fig:exp:sscans:exp1:sel:at},
this improvement reaches 12x to 50x for \athena.

Figure~\ref{fig:exp:sscans:exp1:sel:bq} shows the results
for \bigquery.
The observations are similar, but more pronounced:
Queries with a higher selectivity take longer
for the same reasons as discussed above.
Furthermore, the running time increases with the batch size
due to the larger data volume and higher computational costs
caused by a higher \emph{combined selectivity}.
However, it increases less than the batch size,
thus yielding a considerably higher throughput.
For selectivities smaller than 1\%,
throughput improves by up to 17x,
and for the others, up to 10x.

\textbf{Cost.} The effect of selectivity and batch size on the monetary cost
depends heavily on the pricing model.
For \bigquery, it is a constant \SI{0.011}USD per query batch
for all data points shown in Figure~\ref{fig:exp:sscans:exp1:sel:bq}.
This is due to the fact that
only the number of bytes \emph{of the selected columns} is billed,
which is independent of how many tuples have been selected.
For the above experiments, \SI{4.47}{\gibi\byte} are billed per batch.
The price \emph{per query} hence decreases linearly with the batch size.

\begin{figure}[!h]
    \centering
    \includegraphics[width=0.9\linewidth]{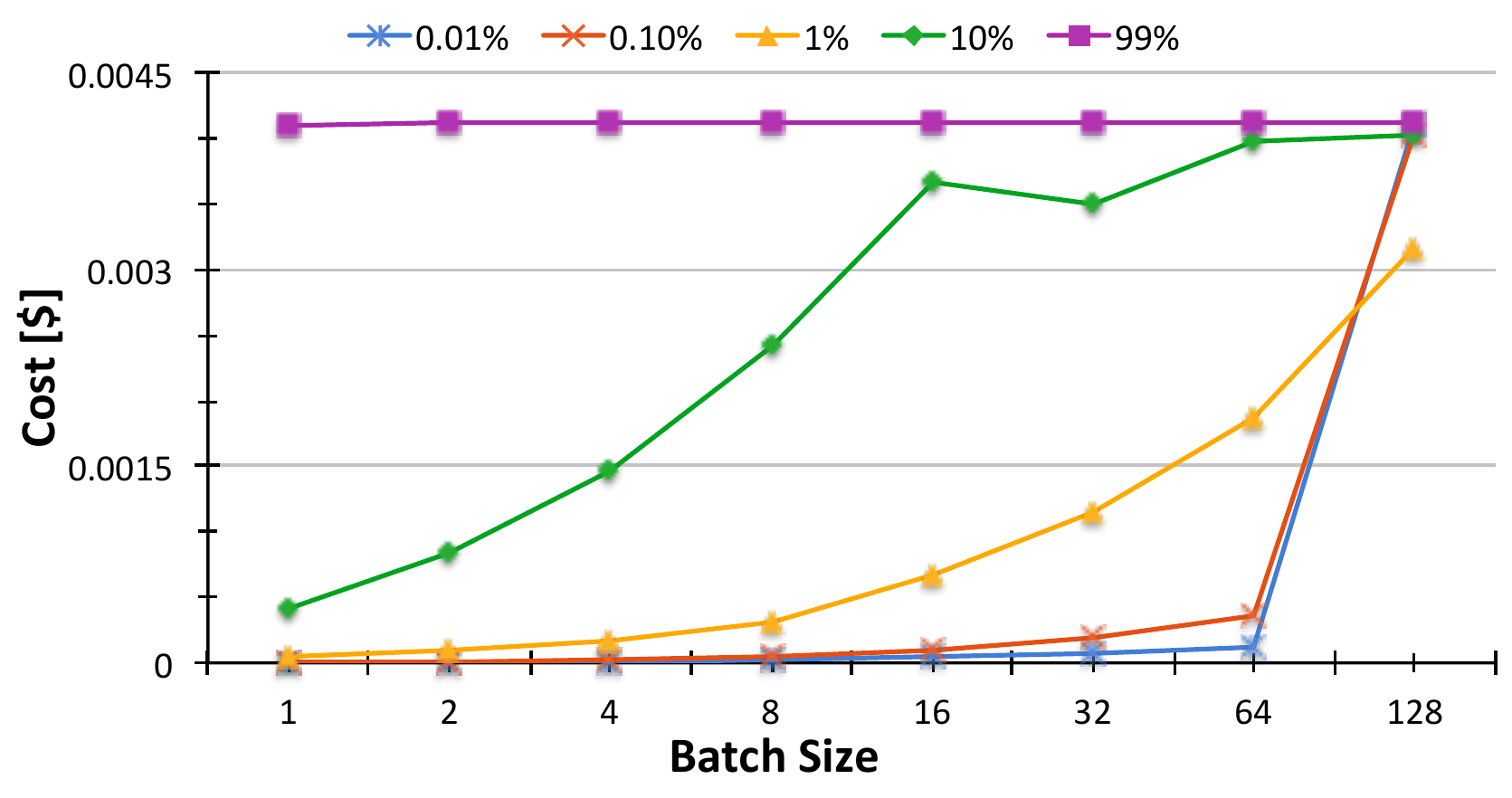}
    \caption{Shared scan query cost in \athena.}
    \label{fig:exp:sscans:exp1:cost:at}
\end{figure}

In \athena, selectivity does affect the monetary costs.
Figure~\ref{fig:exp:sscans:exp1:cost:at} shows how.
Similarly to the discussions about running time,
the cost increases with increasing combined selectivity
of the queries in the batch.
However, unlike above,
the monetary costs do not increase beyond some constant,
namely the cost of reading the entire column.
This corresponds to the constant cost
of the queries with a selectivity of 99\%.
These observations match exactly what the pricing model would suggest,
namely that we pay for the number of bytes read from the storage layer,
which increase with the number of selected tuples
up to the point where all tuples are read.

As a side note, the cost of queries with selectivities of 0.1\% and 0.01\%
jumps to the maximum cost for batches of 128 queries.
These configurations correspond to the unexplainable behaviour
in terms of running time discussed above.
The assumed bug hence also affects monetary costs,
which raises questions about whether the pricing model is fair:
Should users pay more for suboptimal behaviour of the \qaasl system?
This discussion is out of the scope in the paper,
so it is not pursued further.

From the perspective of a single query,
the monetary savings depend on the degree of sharing:
Few queries with low selectivities might not overlap any tuples
and thus cost the same as if executed in isolation,
but for big batches and high selectivites,
the per-query cost may be divided by the batch size.

\subsubsection{Computing the \qset attribute}
\label{sec:exp:sscan:pred_idx}

We now quantify the impact of index predicate evaluation.
To that aim, we generate batches of selection-only queries
using predicates on three different attributes of the LINEITEM table
(l\_discount, l\_quantity, and l\_shipdate).
We compare three approaches how the \qset column is computed:
(1) linear predicate evaluation,
(2) indexed predicate evaluation where only one attribute is indexed
(which corresponds to what dedicated shared execution systems
from prior work~\cite{Unterbrunner:2009:PPU:1687627.1687707} do),
and (3) indexed predicate evaluation where all attributes are indexed.
To show the impact of predicate evaluation,
we do not perform the pre-filtering optimization
described in Section~\ref{sec:shared_scan}.

\begin{figure*}[ht]
    \centering
    \begin{subfigure}{.5\textwidth}
      \centering
      \includegraphics[width=0.9\linewidth]{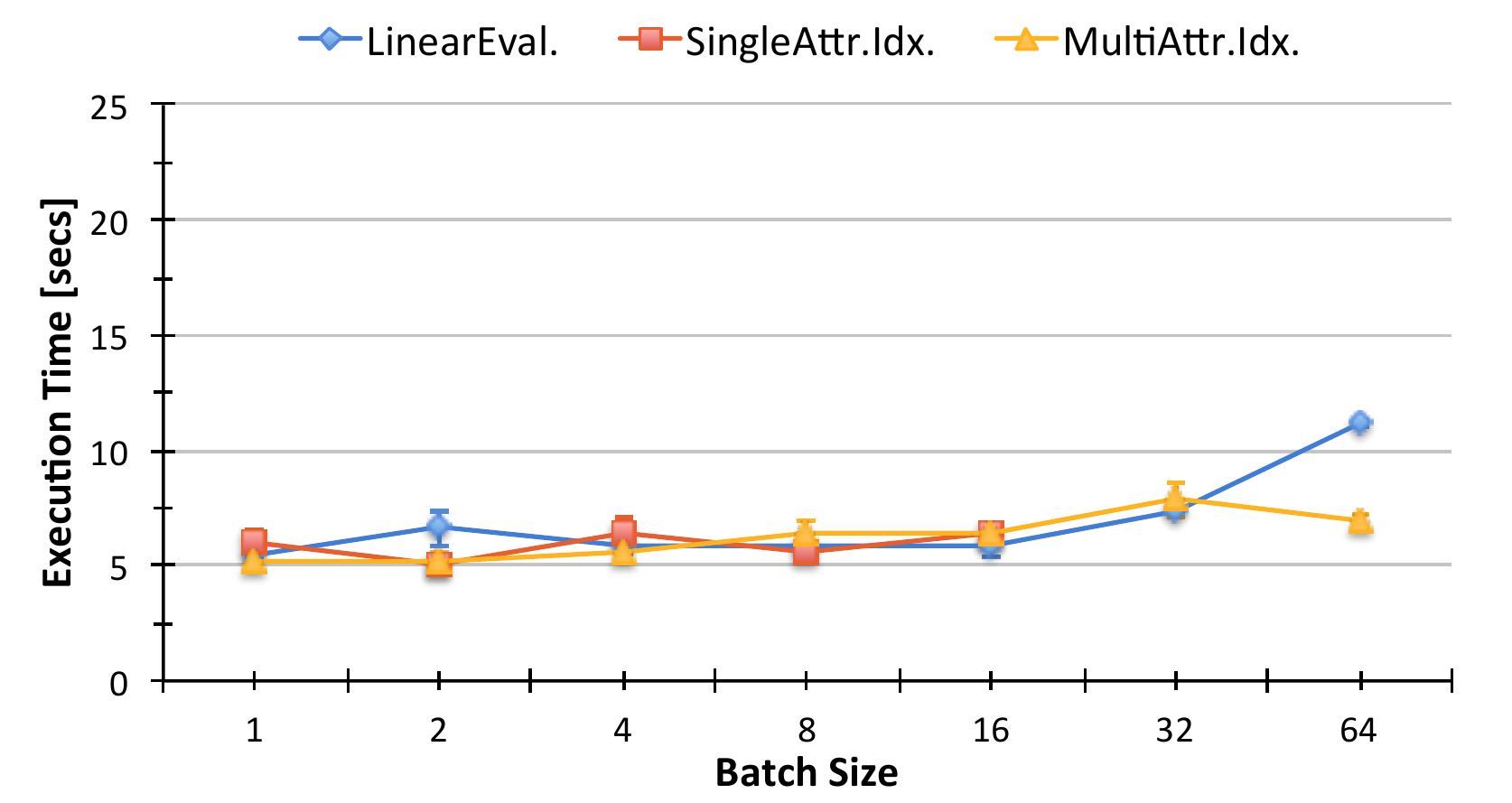}
      \caption{\athena}
      \label{fig:exp:sscans:exp2:at}
    \end{subfigure}%
    \begin{subfigure}{.5\textwidth}
      \centering
      \includegraphics[width=0.9\linewidth]{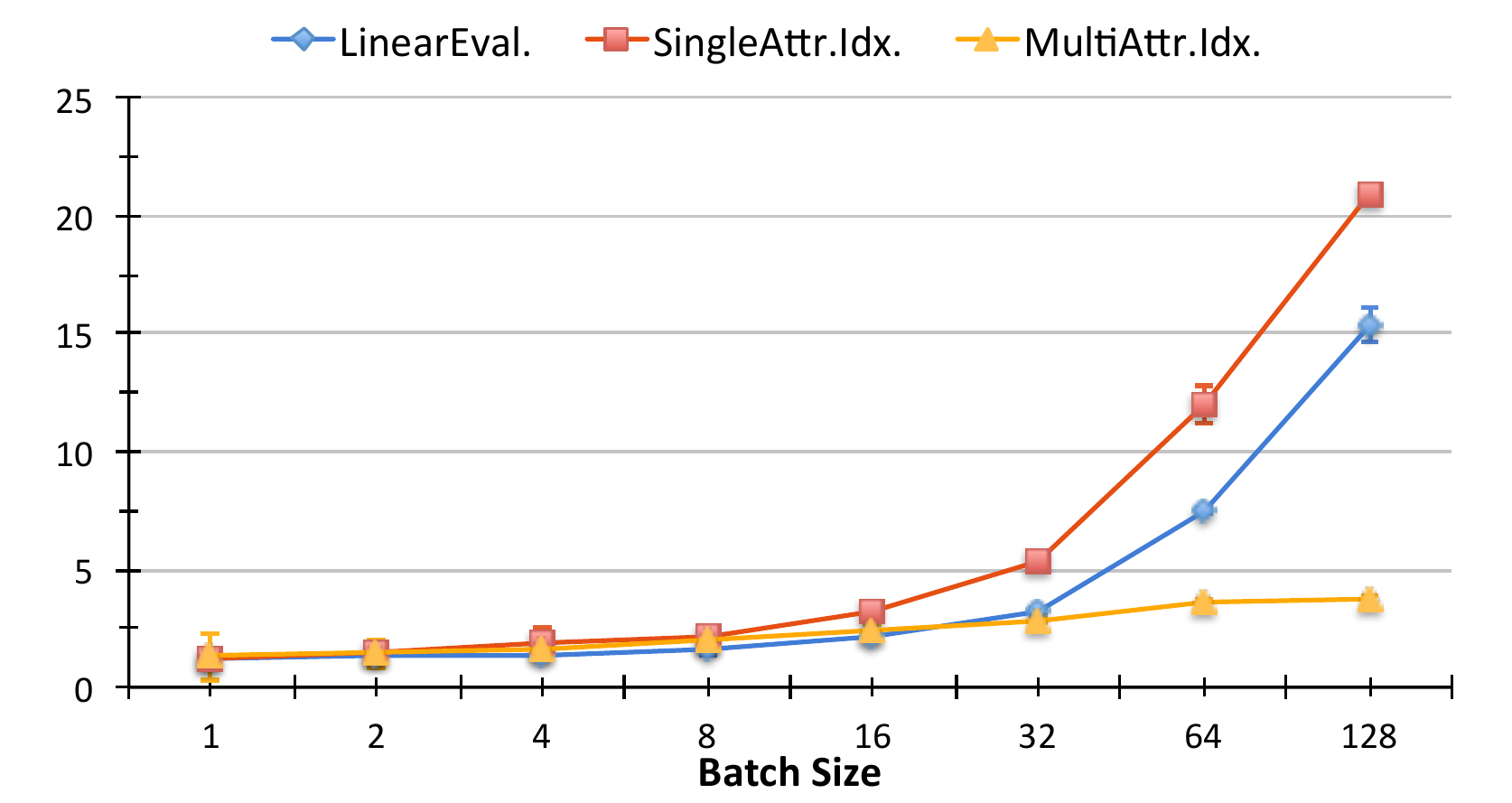}
      \caption{\bigquery}
      \label{fig:exp:sscans:exp2:bq}
    \end{subfigure}
    \caption{Execution time of the shared scan.}
    \label{fig:exp:sscans:exp2:sel}
\end{figure*}

The results for \athena  are shown in Figure~\ref{fig:exp:sscans:exp2:at}. These
results contain a lot of variation for smaller batch sizes, so there is no
clear advantage among the different approaches. However, when batching many
queries together, multi-attribute indexing does pay-off compared
to linearly checking each predicate. When grouping larger number of queries
together, some of the generated queries do
not run and only a generic error is obtained without further explanation or
suggestion indicating what is happening. This might be related to the final size of the
generated \sql queries which in some cases are almost as big as the maximum
allowed limit size, \SI{256}{\kibi\byte}.

Figure~\ref{fig:exp:sscans:exp2:bq} shows the execution time for these different
approaches using \bigquery. We observe that using predicate indexing on a single
attribute does not improve the query execution time because the execution
time is still dominated by the linear predicate evaluation of the other
attributes. Thus, although the first attribute is logarithmic in the number of
queries, the remaining number of comparisons is still linear.  However,
multi-dimensional predicate indexing helps in keeping the number of comparison
logarithmic in the number of queries. This, in combination with the constant
input size, results in an almost constant execution time. 

\subsection{TPC-H workload}
\label{sec:exp:tpch}

We now evaluate the impact of our approach
on end-to-end query performance and monetary cost
on a complex workload derived from TPC-H~\cite{TPCH2018},
a standard database benchmark for decision support queries.

\subsubsection{Workload definition}

We define the workload to consist of 128 instances
of each of the 22 queries defined by the standard,
each with query parameters drawn independently as per the specifications.
We use scale factor 100, which requires \SI{27}{\gibi\byte} in \athena
and \SI{107}{\gibi\byte} in \bigquery.
Unlike the official benchmark, we assume that
the $22 \cdot 128$ queries are ready for execution at once
such that they can be executed jointly.
This mirrors interactive search systems where a search request is translated
into hundreds of parameterized queries for different search attributes.

We show different ways to produce an execution plan for the workload.  One would
expect that a single logical plan for the entire workload is most efficient
because all available sharing opportunities can be exploited.  However, on the
systems we are using, this does not hold due to practical limitations.  Thus, we
show two different alternatives: (1) producing a single logical plan in the form
of a DAG for the entire workload as described in Section~\ref{sec:shared_plans}
(which needs to be executed as several tree-structured plans as explained in
Section~\ref{sec:shared_qplans}) or (2) splitting the workload into one logical
plan for each of the 22 query templates such each batch consists of queries of
the same form.  We concentrate on the latter approach first and give performance
numbers of the other approach later.

For both approaches, we manually produce shared query plans
for the entire workload as described in Section~\ref{sec:sw}
and translate them back to SQL as described in Section~\ref{sec:rewriting}.
We adapted the TPC-H query generator such that it generates
these SQL statements for batches of a configurable number of queries
(while respecting how the query parameters are drawn).
Unlike previous work~\cite{mqjoin,Makreshanski2017}, we preserve the full semantics of TPC-H queries.

\subsubsection{Impact of batch size}
\label{sec:exp:tpch:bsz}

\begin{figure*}[tb]
    \centering
    \begin{subfigure}[b]{\textwidth}
        \includegraphics[width=0.9\textwidth]{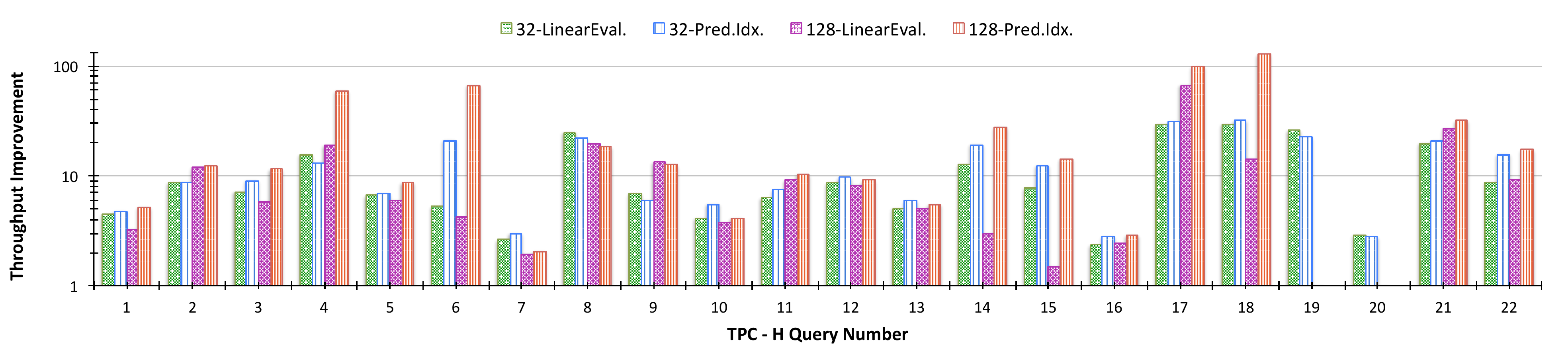}
        \centering
        \caption{\athena}
        \label{fig:in:tpch:throughput:batchsz:at}
    \end{subfigure}
    \begin{subfigure}[b]{\textwidth}
        \includegraphics[width=0.9\textwidth]{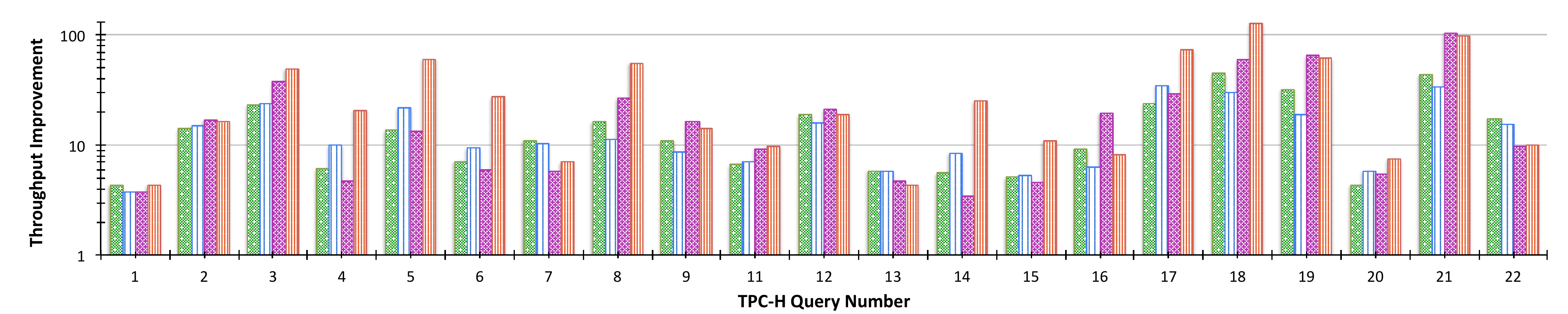}
        \centering
        \caption{\bigquery}
        \label{fig:in:tpch:throughput:batchsz:bq}
    \end{subfigure}
    \caption{TPC-H throughput improvement of shared execution over query-at-a-time.}
    \label{fig:in:tpch:throughput:batchsz}
\end{figure*}

Figure~\ref{fig:in:tpch:throughput:batchsz}
shows throughput improvements thanks to our approach
over the traditional \emph{query-at-a-time} execution,
which consists of running each query independently one after the other.
We execute the workload in batches of both 32 or 128 queries.
While larger batches usually yield a better throughput,
\athena cannot execute all queries at the largest batch size,
so we show the numbers of batch size 32,
which is the largest batch size that works for all queries on both systems.

The upper plot shows the throughput improvement
for \athena for different batch sizes, with indexed predicate evaluation and
without it, compared to executing each query independently. For executing some queries, using a large batch size is actually
not beneficial, e.g., Queries 7 and 10, because replicating the tuples
of the final result set for the final aggregation is compute-bound when a
large number of queries are involved.

The lower plot shows the results for \bigquery,
which are similar to the ones obtained from \athena, except for Query 22.  This
query does not benefit from a larger batch size as \athena does. The reason is
the substring comparisons predicates that are linearly evaluated making it
compute-bound in \bigquery for large batch sizes.  TPC-H Query 10 cannot be run
on \bigquery because it requires sorting on a computed column. Doing so for a
single query does not become memory-bound and \bigquery completes it
successfully.  However, for batches of queries, the order-by operation has to be
carried out for the union of all queries output results which is not
supported by \bigquery.  The sorting operator for large inputs is not available
by design~\cite{resources_exceeded_bq}. 

In general, we can say that a bigger batch of queries improves the overall
throughput if predicate indexing helps in making queries remain disk-bound
(e.g., Queries 4, 6, 17, and 18). If a shared aggregation is needed over a
large input, replicating tuples for the queries in the batch dominates the query
execution.

\subsubsection{Predicate indexing}
\label{sec:exp:tpch:pred_idx}

All queries shown use predicate indexing wherever possible. There are queries,
however, that contain predicate types we cannot currently index  (Queries 7, 9,
12, 13, 16, 19, and 22) as previously explained.

Figures~\ref{fig:in:tpch:throughput:batchsz} show the throughput improvement
when doing a \textit{linear evaluation} of all predicates, and when using
\textit{indexed predicates}, for computing the \qset column for a batch of
TPC-H queries.

In general, queries benefit the most from predicate indexing if it is applied
when scanning the largest relations. For instance, in Query 3 we are able to
index the predicates used over the three largest relations (CUSTOMERS, ORDERS,
and LINEITEM). However, there is not a bigger improvement because it
still requires replicating tuples for each query in the batch for the final
aggregation.

The predicates of Queries 4 and 5 are over the second largest relation (ORDERS).
For these queries, we do early tuple replication before carrying out the
joins to avoid having to replicate even more tuples resulting from
the join. After that, queries can continue as regular non-batched statements.

Query 6 presents the biggest improvement. It basically consists on scanning the
largest relation (LINEITEM) where the different predicates are on multiple
attributes but with rather small attribute ranges. This makes each predicate
index structure shallow, which results in a lower total number of
comparisons for generating the \qset column. This is the best scenario for using
indexed predicates.

Executing Query 10 does not work on \bigquery as discussed above. In \athena, this query can be
successfully executed and its runtime improvement comes from using indexed
predicates and from doing an early tuple replication for avoiding to replicate
even more tuples after performing the query joins.

Queries 14, 15, 17, and 18 are also improved by using indexed predicates on their
large relations due to the fact that computing the \qset column can
dominate the overall execution. In general, queries using indexed predicates
over large relations benefit the most from it.

    \begin{table*}[tb]
        \centering
        \resizebox{\textwidth}{!}{
            \begin{tabular}{@{}rrrrrrrrrrrrrrrrrrrrrrr@{}}
                \toprule
                \textbf{TPC-H Query}&1&2&3&4&5&6&7&8&9&10&11&12&13&14&15&16&17&18&19&20&21&22\\ 
                \midrule
    
                \textbf{BatchSz.}
                & 128  & 128  & 128  & 128 & 128 &128 & 32  &64 & 128 &64& 128 &64 & 32
                & 128 & 128 & 64  &128& 128 &32 & 64  &128& 128\\
    
                \textbf{Pred.Idx}
                & \cmark  & \xmark & \cmark  &\cmark & \cmark  &\cmark & \cmark &\xmark
                & \xmark  &\cmark& \xmark  &\xmark & \cmark & \cmark &\cmark & \xmark
                &\cmark& \cmark &\xmark & \cmark &\cmark& \cmark\\
    
                \midrule
    
                \textbf{SINGLE} Qry.Exec. [s]
                &5.78    &14.24   &7.43   &7.13&  11.49&  5.51&  18.84&  10.86
                &14.58   &7.85    &6.24   &6.68&  8.68 &  6.28&  6.76 &  3.11
                &16.11   &12.09   &5.89   &7.20&  40.50&  5.03\\
                \midrule
    
                Runtime [s]
                &142.7   & 147.9    &83.13  & 15.41   & 169.8  &10.61
                &808.3   & 57.29    &137.4  & 171.3 & 75.32    &77.41     
                &187.6    & 29.06   &60.7   & 131.8   & 21.05   & 12.23  & 31.20   &322.0
                &163.1   & 37.28\\
    
                Thr.Imp.
                &5.179    &11.91  & 11.43&   59.24& 8.658 & 66.48
                &2.984    &27.20  & 13.26&   5.864 &9.087  & 9.880   &5.917 &27.64
                &14.24    &2.455   &97.90&  126.4& 25.82& 2.862   &31.76&17.26 \\
    
                \midrule
                SharedExec. [\$]
                &0.027&  0.005&  0.031&  0.026&  0.029&  0.028&  0.147&  0.077
                &0.041&  0.049&  0.006&  0.073&  0.046&  0.033&  0.063&  0.003
                &0.035&  0.017&  0.477&  0.133&  0.081&  0.004\\
    
                Query-at-a-time [\$] 
                &3.417  &0.630&  3.936&  3.388&  3.697&  3.593
                &4.705  &4.955&  5.259&  3.159&  0.711&  4.654&  1.463 & 4.166
                &8.060  &0.178&  4.434&  2.199&  3.819&  4.269&  10.35&  0.505 \\
    
                \bottomrule
            \end{tabular}
        }
        \caption{Best configurations for \athena}
        \label{tab:bconfig:at}
    \end{table*}

    \begin{table*}[tb]
        \centering
        \resizebox{\textwidth}{!}{
            \begin{tabular}{@{}rrrrrrrrrrrrrrrrrrrrrrr@{}}
                \toprule
                \textbf{TPC-H Query}&1&2&3&4&5&6&7&8&9&10&11&12&13&14&15&16&17&18&19&20&21&22\\ 
                \midrule
                \textbf{BatchSz.}
                &32 &   64  &128    &128    &128    &128 &32    &128    &128 &1  &   128 &128    &64
                &128&   128 &128    &128    &128    &128 &128   &128    &32 \\
    
                \textbf{Pred.Idx}
                &\xmark &\xmark &\cmark &\cmark &\cmark 
                &\cmark &\xmark& \cmark &\xmark &\xmark 
                &\cmark &\cmark &\xmark &\cmark &\cmark
                &\xmark &\cmark &\cmark &\xmark &\cmark 
                &\cmark &\xmark \\
    
                \midrule
                \textbf{SINGLE} Qry.Exec. [s]
                &5.53   &11.77&  51.14&  4.90 &  12.37& 1.97&  8.55 &   36.52&   9.20
                &9.90   &4.96 &  3.49 &  10.66&  2.68 & 4.02&  18.78&   10.03
                &13.54  &2.03 &  11.99&  82.89&  10.00\\
    
                \midrule
    
                Runtime [s]
                &145.7&    81.17 &   133.0&    30.17&    26.08&    9.166&    95.88&    83.85
                &71.17&    1267.2&    64.7&    23.61&    205.7&    13.65&    47.77&    180.3
                &17.68&    13.53 &   4.026&    205.3&    107.6&    69.60\\
    
                Thr.Imp.
                &4.318   & 17.85&    49.21&    20.79&    60.68
                &27.48   & 10.97&    55.73&    16.43&    1    &   9.796 &   18.91
                &6.000   & 25.16&    10.77&    19.57&    72.59&   128.03&   64.69
                &7.473   & 98.55&    17.36\\
    
                \midrule
                SharedExec. [\$]
                &0.502&  0.029&  0.110&  0.083&  0.105&  0.087&  0.485&  0.129
                &0.155&  12.94&  0.012&  0.117&  0.092&  0.090&  0.088&  0.011
                &0.068&  0.067&  0.146&  0.100&  0.095&  0.031\\
    
                Query-at-a-time [\$] 
                &16.07    &1.84    &14.14    &10.69  & 13.42 &   11.18
                &15.52    &16.57   &19.79    &12.94  &  1.50 &   14.98&    5.88&
                11.53 &11.22    &1.35    &8.70     &8.63   & 18.68 &   12.83&
                12.16& 0.99\\
    
                \bottomrule
            \end{tabular}
        }
        \caption{Best configurations for \bigquery}
        \label{tab:bconfig:bq}
    \end{table*}

\subsubsection{TPC-H cost analysis}
\label{sec:exp:tpch:cost_analysis}

For the TPC-H workload, varying the number of queries grouped does not
increase the monetary cost significantly, i.e.,
executing a single query is as expensive as executing a group of queries sharing
the same execution plan.

Tables~\ref{tab:bconfig:at} and ~\ref{tab:bconfig:bq} show the best
configuration (batch size, and \qset attribute computation method), query
execution time, and cost for obtaining the fastest execution time of the
workload. Although throughput increases with the batch size, individual query
latency also increases as they have to be grouped. The best execution
time is not always achieved with the largest batch size. For instance, executing
Query 7 with batch sizes of 32 is faster than executing a batch of 128
queries in both systems, but it is also 4x times more expensive, i.e.,
executing 4 times a batch of 32 queries.

The workload execution with sharing yields a lower execution cost
compared to executing queries one at the time.  For \athena, running this
workload  without sharing costs 81.54 USD. With sharing using large batches
it costs 0.759 USD, i.e., it is 107x cheaper. This cost saving relates directly
to the batch size of 128 queries used. Further monetary cost improvements could
be achieved if larger batch sizes were used. For \bigquery, running a complete
TPC-H run without sharing costs 240.59 USD and with sharing using large batches
it costs 14.72 USD, including  Query 10 which we cannot optimize, i.e., it is
16x cheaper. If Query 10 is not taken into account, it is 128x cheaper.

\subsubsection{Global shared plan}
\label{sec:exp:tpch:global_splan}

We now show how to execute the workload using a single logical plan.
This has a higher sharing potential than executing them grouped by
type as in the previous experiments.
We thus produce a single logical plan in the form
of a DAG for the entire workload as described in Section~\ref{sec:shared_plans}.
Note that this global logical plan produces 22 results,
one for each of TPC-H queries.
We transform this plan into several tree-structured plans as explained in
Section~\ref{sec:shared_qplans}. 
Since a cost-based optimizer is out of the scope of this paper,
we do the transformation manually.
As general strategy, we materialized the joins with large results used by multiple
    queries, and recompute the ones with smaller results.

Furthermore, we do not include the \qset attribute
in the materialized intermediate results
because recomputing it would not incur in extra monetary costs, but reading it would.

We carry on this experiment only
in \bigquery as \athena does not support reusing intermediate results in
columnar format (it only supports row-oriented text format for intermediate
results) which would make this approach extremely inefficient and expensive.
We compare our two approaches for describing the limitations of the current
implementation.

\begin{figure}[b]
    \centering
        \includegraphics[width=0.8\linewidth]{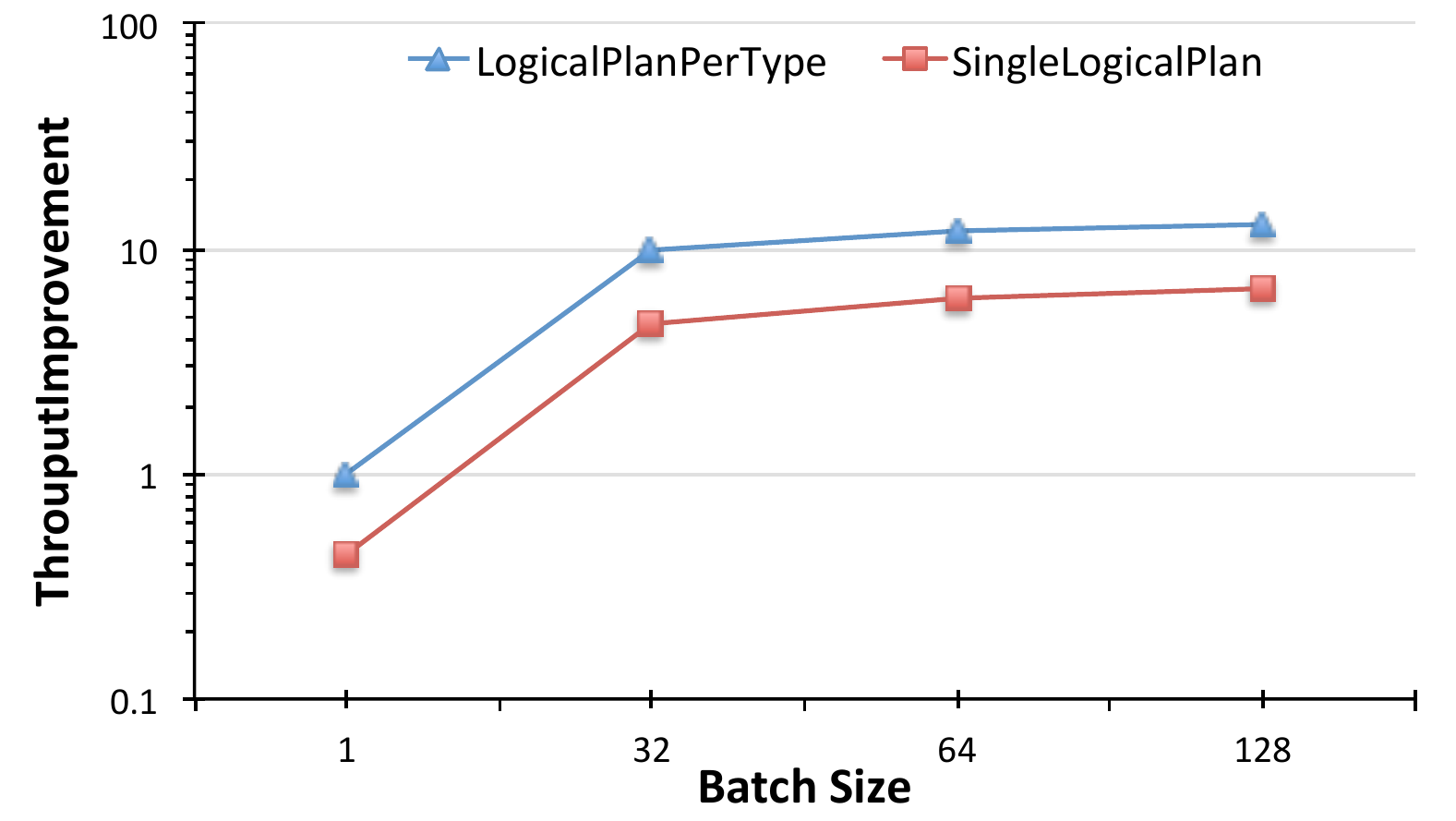}
        \caption{Throughput improvement of execution modes.}
        \label{fig:exp:th_imp_over_qat}
\end{figure}

Figure \ref{fig:exp:th_imp_over_qat} shows the throughput improvement of both
approaches.  The lower throughput improvement achieved by the global shared plan
is due to (1) the materialization step of common intermediate results and (2)
queries accessing more data than required because the materialized common
results might be larger than needed for a given query.  In spite of this, there
is a throughput improvement once there are enough queries to group and execute
afterwards.  

\begin{figure}[!hb]
    \centering
        \includegraphics[width=0.9\linewidth]{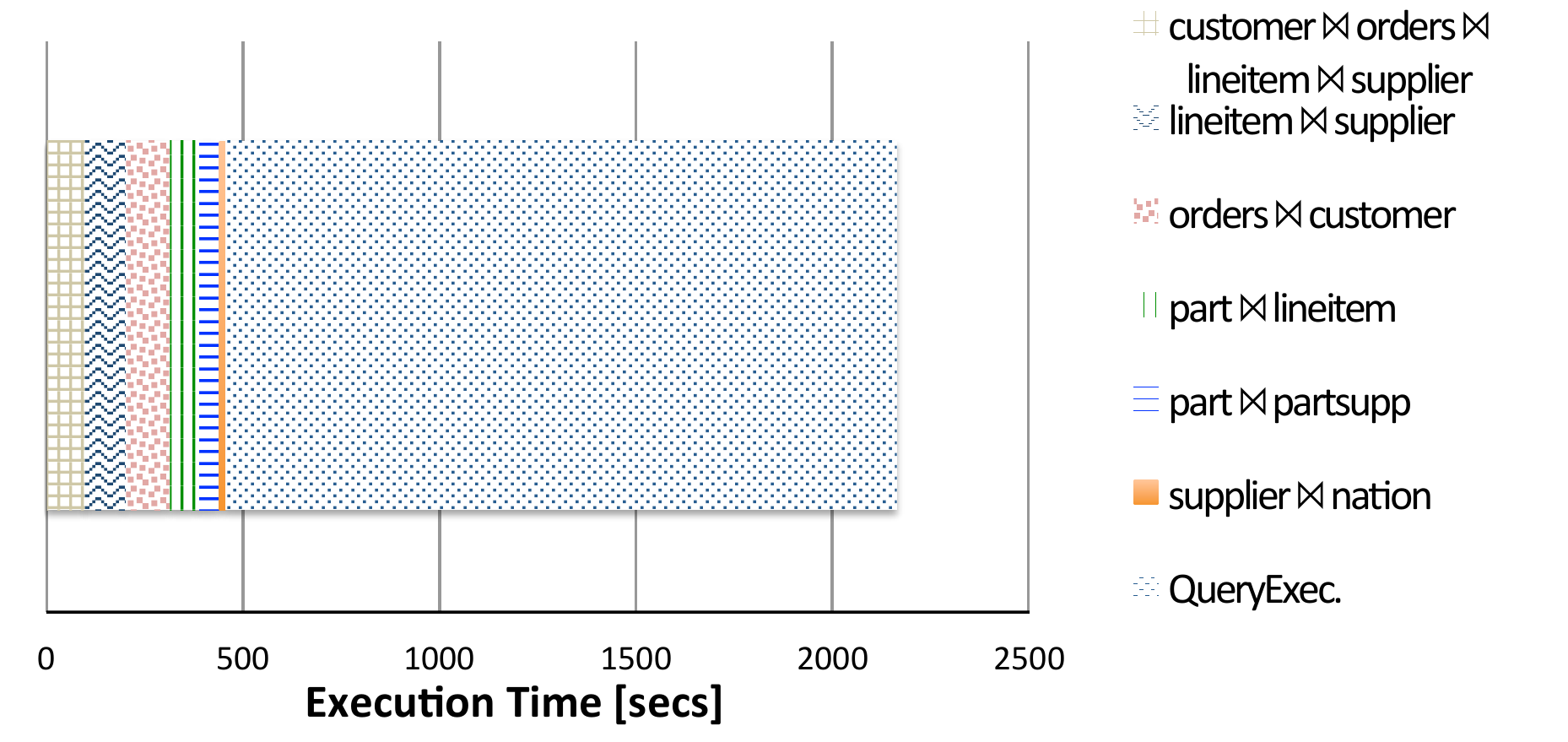}
        \caption{Execution time breakdown}
        \label{fig:exp:exec_time_breakdown}
\end{figure}

Figure \ref{fig:exp:exec_time_breakdown} shows how much of the overall execution
time goes into materializing intermediate results when using groups of 32
queries each. The materialization time accounts for 21\% of the time of
executing a workload of 32 x 22 queries. For this workload, it results in a 5x
and 9.7x throughput and cost improvement, respectively. The absolute time of
materializing intermediate results does not go down with more queries being
grouped because with just a few queries of different types we end up requiring
most of the data from the base tables.

\section{Conclusions}
\label{sec:conclusions}

In this paper, we apply shared-workload techniques at the \sql level for
improving the throughput of \qaasl systems without incurring in additional
query execution costs. Our approach is based on query rewriting for grouping
multiple queries together into a single query to be executed in one go. This
results in a significant reduction of the aggregated data access done by the
shared execution compared to executing queries independently.



We presented a cost and runtime evaluation of the shared operator driving data access costs. 
Our experimental study using the TPC-H benchmark confirmed the benefits of our
query rewrite approach. Using a shared execution approach reduces significantly
the execution costs. For \athena, we are able to make it 107x cheaper and for
\bigquery, 16x cheaper taking into account Query 10 which we cannot execute,
but 128x if it is not taken into account. Moreover, when having queries that do
not share their entire execution plan, i.e., using a single global plan, we
demonstrated that it is possible to improve throughput and obtain a 10x cost
reduction in \bigquery.


There are multiple ways to extend our work. The first is
to implement a full \sql-to-\sql translation layer to encapsulate the proposed
per-operator rewrites.  Another one is to incorporate the initial work on
building a cost-based optimizer for shared execution
\cite{Giannikis:2014:SWO:2732279.2732280} as an external component for \qaasl
systems.  Moreover, incorporating different lines of work (e.g., adding
provenance computation \cite{GA09} capabilities) also based on query
rewriting is part of our future work to enhance our system.

\balance
\printbibliography

\end{document}